\documentclass[prb,aip,showpacs,twocolumn,amsmath,amssymb]{revtex4}
\usepackage{colordvi}
\usepackage{epsfig}
\begin{document}
\title{Spin relaxation in $n$-type GaAs quantum wells with
transient spin grating}
\author{M. Q. Weng}
\email{weng@ustc.edu.cn.}
\affiliation{Hefei National Laboratory for Physical Sciences at
  Microscale, University of Science and
Technology of China, Hefei, Anhui, 230026, China}
\affiliation{Department of Physics, University of Science and
Technology of China, Hefei, Anhui, 230026, China}
\altaffiliation{Mailing address.}
\author{M. W. Wu}
\email{mwwu@ustc.edu.cn.}
\affiliation{Hefei National Laboratory for Physical Sciences at
  Microscale, University of Science and
Technology of China, Hefei, Anhui, 230026, China}
\affiliation{Department of Physics, University of Science and
Technology of China, Hefei, Anhui, 230026, China}
\altaffiliation{Mailing address.}
\author{H. L. Cui}
\affiliation{Department of Physics and Engineering Physics, Stevens
  Institute of Technology, Hoboken, NJ 07030, USA}

\date{\today}

\begin{abstract}
By solving the kinetic spin Bloch equations, we study the time
evolution of the transient spin grating, whose spin polarization
varies periodically in real space, confined in (001) GaAs quantum wells.
With this study we can investigate the properties of both the spin transport
 and the spin relaxation at the same time.
The Fourier component of the spin signal  decays
double exponentially with two decay rates
$1/\tau_+$ and $1/\tau_-$.
In  high temperature regime, the average of these
two rates varies with the grating wave-vector
 $q$ quadratically, i.e.,
$(1/\tau_++1/\tau_-)/2=D_sq^2+1/\tilde{\tau}_s$, with
$D_s$ and $\tilde{\tau}_s$ representing the spin diffusion coefficient
and the average of the out-of-plane and the in-plane
spin relaxation times  respectively.
 $\tau_{\pm}$ calculated from our theory are in good agreement with the
  experimental data by Weber {\em et al.} [Phys. Rev. Lett.
{\bf 98}, 076604 (2007)].
By comparing $D_s$ with and without the
electron-electron Coulomb scattering, we calculate the
contribution of Coulomb drag to the spin diffusion coefficient.
With the transient spin grating result, we further reveal the
relations among different characteristic parameters such as spin
diffusion coefficient $D_s$, spin relaxation time $\tau_s$, and
spin injection length $L_s$. We show that in the presence of the Dresselhaus and/or Rashba
spin-orbit coupling, the widely used relation
$L_s=\sqrt{D_s\tau_s}$ is generally inaccurate and can even be
very wrong in some special cases. We present
an accurate way to extract  the steady-state
transport characteristic parameters
from the transient spin grating signals.
\end{abstract}


\maketitle

\section{Introduction}

Recently a lot of efforts have been devoted to the study of  spin
dynamics and spin transport in semiconductor nano-structures in order
to realize the spintronic device.\cite{wolf,spintronics,das,wubook}
In $n$-type zinc-blende semiconductors, electron spins are
randomized by the Dresselhaus and/or
the Rashba  spin-orbit coupling (SOC) which acts as an
effective magnetic field ${\bf h}({\bf k})$ with its direction and
magnitude depending on the electron momentum
${\bf k}$.\cite{dp,dpb,rashba}
In spatial homogeneous system the spin evolution is characterized by the
spin relaxation time  $\tau_s$ which describes the
decay rate of spin polarization; The steady-state spin transport
 is characterized mainly by the spin injection/diffusion length
$L_s$,\cite{schmidt,flatte,zuti,zuti2}
while the transient spin propagation is characterized by spin diffusion
coefficient $D_s$.\cite{miller_1996,weber_2005} The relations
among these three parameters and other parameters
such as momentum relaxation time $\tau_p$, charge diffusion
coefficient $D_c$ and mobility $\mu$ have been actively
discussed.

It is understood that spin relaxation/dephasing is induced by the inhomogeneous
broadening due to the SOC together with the (spin conserving)
scattering.\cite{wu_epjb_2000,wu_jpsj_2001}
The scattering provides a channel to speed up the spin dephasing
but also slows down the spin dephasing by weakening the inhomogeneous
broadening.\cite{weng_prb_2003} The
competing effects of scattering have different results on spin dephasing in
different conditions. In weak SOC or strong scattering regime,
the out-of-plane spin relaxation  time  is
expected to be $1/\tau_s\propto \langle
{\bf h}^2({\bf k})\tau^{\ast}_p\rangle$.  It should be noted
that $\tau^{\ast}_p$ not only includes the contribution from
conventional momentum scattering mechanisms such as electron-impurity
and electron-phonon scattering, but also includes the contribution
from electron-electron Coulomb scattering which does not directly
affect the charge transport
properties.\cite{wu_epjb_2000,glazov_2002,weng_prb_2003}
As for the steady-state spin
injection problem, by assuming that the spin dynamics can be separated
into two independent processes, spin diffusion and spin relaxation,
the spin polarization is expected to decay exponentially along the
injection direction with decay ``rate'', i.e., the spin
injection length, $L_s=\sqrt{D_s\tau_s}$, where
$D_s$ and $\tau_s$ are two phenomenal parameters whose relations with
other properties are yet to be determined.\cite{schmidt,flatte,zuti,zuti2}
However in the presence of the SOC, it has been proved that this assumption is
oversimplified. By solving the  kinetic spin Bloch
equations, it is shown that the effective magnetic field due to the SOC
alone causes the electron spin to process in real space even in diffusive
regime.\cite{weng_prb_2002,weng_jap_2003}
The spin polarization varies in the space as $e^{-x/L_s}\cos(x/L_0)$
instead of simple exponential decay. The spin injection length $L_s$
and the spatial oscillation ``period'' $L_0$ are
obtained by solving the kinetic spin Bloch equations,\cite{cheng_jap_2007}
the spin transport equations which include the contribution of the
SOC\cite{stanescu_07} or  the linear response
theory.\cite{burkov_2004} In the diffusive regime, $L_s$ and
$L_0$ can be expressed by SOC strength and momentum scattering
time.\cite{cheng_jap_2007,stanescu_07,burkov_2004}
As for the spin diffusion coefficient, it was widely assumed to be the
same as the charge diffusion coefficient.
Later it was pointed out that spin Coulomb drag (SCD), caused by the
electron-electron Coulomb scattering, should suppress the relative motion
of electrons with different spins and thus reduce
$D_s$.\cite{amico_2002,amico_2003,jiang_2005,weber_2005}

A direct measurement of spin diffusion
coefficient can be carried out by transient spin grating (TSG)
experiments.\cite{miller_1996,weber_2005,carter_06,weber_07}
With the assumption that spin diffusion and spin relaxation are
independent of each other, the decay rate of TSG was written as
$\Gamma_q=D_sq^2+1/\tau_s$, where $q$ is the wave-vector of the spin
grating.\cite{miller_1996,weber_2005} However, from the lesson of
steady-state spin injection one learns that the spin diffusion and
relaxation are not separable even in the diffusive regime. This is further
justified by the fact that the decay of TSG can be fitted to a
double-exponential form instead of single exponential
one.\cite{weber_07} It is therefore still a question of how to get spin diffusion coefficient
through TSG experiments.

In this article we study the temporal evolution of the TSG by
solving the kinetic spin Bloch equations. This paper is
organized as following: In Sec.\ \ref{kineticequation} we
first set up the  kinetic spin Bloch equations and apply them to
solve the TSG problem. Then we show the analytical solution of simplified
equations and the numerical results of the full
kinetic spin Bloch
equations. In Sec.\ \ref{statictransport} we study the relations among
the spin relaxation
time $\tau_s$ of a spatially homogeneous system, the spin injection
length $L_s$ and the spatial
oscillation period $L_0$ of the steady-state spin transport as well as $D_s$ of
the transient spin transport. We conclude in Sec.\ IV.

\section{Kinetic Spin Bloch Equations}
\label{kineticequation}
The system we study is the electron gas confined in a $(001)$
GaAs quantum well (QW) with width $a$ grown along the $z$-axis. We
assume that the well width is narrow enough so that only the lowest subband is
occupied. With the help of the nonequilibrium Green function method,\cite{haug}
one can write down the kinetic spin Bloch
equations\cite{wubook,weng_jap_2003,weng_prb_2002}
by using gradient expansion and generalized Kandanoff-Baym ansatz:
\begin{widetext}
\begin{eqnarray}
&&{\partial\rho_{{\bf k}}(x,t) \over \partial t} +
    eE(x){\partial\rho_{{\bf k}}(x,t)\over\partial k_x}-
    {k_x\over m}
    {\partial\rho_{{\bf k}}(x,t)\over\partial x}
- i[(g\mu_B{\bf B}
+{\bf h}({\bf k}))\cdot
\mbox{\boldmath $\sigma$}/2
+\varepsilon_{\mathtt{HF}}(x,{\bf k},t),
\rho_{{\bf k}}(x,t)]\nonumber\\
&&=\left.{\partial \rho_{{\bf k}}(x,t)\over
  \partial t}\right|_{\mathtt{s}}.
\label{eq:kinetic}
\end{eqnarray}
\end{widetext}
Here we assume that the transport direction is along the $x$-axis.
$\rho(x,{\bf k},t)$ is the  density matrix
whose diagonal elements $f_{{\bf k}\sigma}(x,t)$ represent
the electron distribution functions  with spin $\sigma$($=\pm1/2$) and momentum
${\bf k}=(k_x,k_y)$ at position $x$. The off-diagonal elements
stand for the spin correlations between spin-up and
-down electrons.
The second and third terms of Eq.\ (\ref{eq:kinetic})
correspond to the drift driven by the electric field
$E(x)$, determined by the Poisson equation, and the diffusion of
electrons, respectively. The fourth term describes the spin precession
around the total magnetic field which is composed of the external
magnetic field ${\bf B}$, the effective
magnetic field ${\bf h}({\bf k})$ due to the
SOC as well as the one from the Hartree-Fock term of the electron-electron
Coulomb interaction $\varepsilon_{\mathtt{HF}}(x,{\bf k},t)$.
${\bf h}({\bf k})$ contains the Dresselhaus and the Rashba
terms:\cite{dp,dpb,rashba}
\begin{widetext}
\begin{eqnarray}
{\bf h}({\bf k})&=&\beta
(-k_x\cos 2\theta+k_y\sin 2\theta,
k_x\sin 2\theta+k_y\cos 2\theta,0)
\nonumber\\&&
+\gamma({k_x^2-k_y^2\over 2}\sin 2\theta
+k_xk_y\cos 2\theta)(k_y,-k_x,0)
+\alpha(k_y,-k_x,0)\ ,
\label{eq:hk}
\end{eqnarray}
\end{widetext}
where $\theta$ is the angle between $x$-axis (the spin
injection/diffusion direction) and the
$(100)$ crystal axis.\cite{cheng_prb_2007} $\beta=\gamma \pi^2/a^2$ with
$\gamma$ being the Dresselhaus coefficient.\cite{dp,dpb}
$\alpha$ represents the Rashba parameter which depends on the electric field
along the  growth direction of the QW.
The scattering term $\left.{\partial \rho_{{\bf k}}(x,t)\over \partial
  t}\right|_{\mathtt{s}}$ includes all the scattering, i.e.,
the electron-impurity, the electron-phonon and most importantly
the electron-electron Coulomb scattering.
It is noted that in our calculation the
  electron-electron interaction is treated beyond the Hartree-Fock
  approximation.
The expressions for the Hartree-Fock
and the scattering terms are given in detail in
Ref.\ \onlinecite{weng_prb_2004b}.

The kinetic spin Bloch equations describe the spin dynamics in the presence of
drift, diffusion and spin precession. By choosing some specified
initial and boundary conditions, one can obtain the evolution of the spin
signal in time and real space by solving these equations for different
systems. In order to study the TSG, the initial spin
polarization of the electrons is chosen to be a sinusoidal wave along
the $x$-direction
$P(x)=(N_{\uparrow}(x)-N_{\downarrow}(x))/(N_{\uparrow}(x)+N_{\downarrow}(x))
=P_0\sin(x/L)$ but uniform along the $y$-axis, where $L=2\pi/q$ is the spatial
period. Using periodical boundary condition, one only needs to study the
dynamics in one period of the space regime.

Unless one makes some simplifications, the  kinetic spin
Bloch equations are
too complicated to be solved analytically. In this paper, we first
present an analytical solution in the diffusive regime using simplified
equations. This solution can only provide an
intuitive vision of the TSG dynamics.
We then present the numerical
solution of the full kinetic spin Bloch equations.

\subsection{Simplified Solution}
\label{simp}

By neglecting the Hartree-Fock term, the inelastic scattering such as the
electron-phonon and the electron-electron Coulomb
scatterings and the coupling to
the Poisson equation,  one is able to rewrite Eq.\ (\ref{eq:kinetic})
in diffusive regime
when the scattering is strong enough, by
using a similar method for calculating the
spin relaxation  as in Refs.\ \onlinecite{dp,dpb},
\begin{widetext}
\begin{eqnarray}
&&{\partial\bar{{\bf S}}(q,t)\over \partial t}
  +Dq^2
  \bar{{\bf S}}(q,t)
  +iq\bar{{\bf h}}\times\bar{{\bf S}}(q,t)\nonumber\\
&&+
{1\over 2}\left(
    \begin{array}{ccc}
      (\alpha^2 +\hat{\beta}^2- 2\alpha\hat{\beta}\sin
      2\theta){\langle k^2\tau_1\rangle}\atop
      +{\langle(\gamma k^3)^2\tau_3\rangle}/{16}
      &
      2\alpha\hat{\beta}{\langle k^2\tau_1\rangle}\cos 2\theta
      & 0\\
      2\alpha\hat{\beta}{\langle k^2\tau_1\rangle}\cos 2\theta &
      (\alpha^2 +\hat{\beta}^2+2\alpha\hat{\beta}\sin
      2\theta){\langle k^2\tau_1\rangle}\atop
      +{\langle(\gamma k^3)^2\tau_3\rangle}/{16}
      & 0\\
      0&0 &
      {2(\alpha^2+\hat{\beta}^2) \langle k^2\tau_1\rangle
        \atop
      +{\langle(\gamma k^3)^2\tau_3\rangle/8}
    }
    \end{array}
  \right)\bar{{\bf S}}(q,t)
  =0\ .
  \label{eq:2}
\end{eqnarray}
\end{widetext}
Here $\bar{{\bf S}}(q,t)=\int d^2{\bf k}
\mbox{Tr}\{\rho(q,{\bf k},t)
\mbox{\boldmath{$\sigma$}}\}{d\theta}$
is the Fourier component of the spin density function.
The second term  represents the diffusion term with the
diffusion constant $D=\langle k^2\tau_1/2m^2\rangle$.
$1/\tau_l=\int_0^{2\pi}{1\over\tau(k,\theta)}\cos
(l\theta)d\theta/2\pi$ with $\tau(k,\theta)$ being the momentum
relaxation time due to the electron-impurity scattering.
The third term is the spin rotation caused by the non-vanishing net
effective magnetic field
\begin{equation}
\bar{{\bf h}}=\langle k^2\tau_1/m\rangle(-\hat{\beta}\cos 2\theta,
\hat{\beta}\sin 2\theta-\alpha,0)
\label{eq:hq}
\end{equation}
due to the DP effect and the diffusion.
The last term is the spin relaxation caused by the DP effect
and the spin conserving scattering.
For a system not far away from the equilibrium,
$\langle\cdots\rangle=\int \cdots \partial
  f(\varepsilon_{{\bf k}})/\partial \varepsilon_{{\bf k}}
d^2{\bf k} /\int \partial
  f(\varepsilon_{{\bf k}})/\partial \varepsilon_{{\bf k}}
d^2{\bf k}$ with $f(\varepsilon)$ being the Fermi  distribution function.
It is noted that similar results at low temperature have been
obtained by different approach recently.\cite{stanescu_07}

In the presence of both the Dresselhaus and the Rashba terms, the spin
relaxation rates are highly anisotropic.
In the weak SOC regime, the in-plane spin relaxation rates are
characterized by
two decay rates
$\langle(\alpha\pm\hat{\beta})^2
k^2\tau_1\rangle/2+\gamma ^2\langle k^6\tau_3\rangle/32$,
corresponding to the spin relaxation along the characterized
directions ($110$) and ($1\bar{1}0$) axises respectively.
While the out-of-plane relaxation rate is
the sum of these two rates
$1/\tau_s=\langle(\alpha^2+\hat{\beta}^2)
k^2\tau_1\rangle+\gamma ^2\langle k^6\tau_3\rangle/16$.
Note that, for the quasi-two-dimension system, the presence of the
 cubic Dresselhaus term modifies the coefficient of linear
 Dresselhaus term $\beta$ to be $\hat{\beta}=\beta-\gamma
 k^2/4$. This relation together with Eq.\ (\ref{eq:2}) are
 briefly derived in the Appendix\ A.
Without the spin rotation term, the evolution of the TSG is characterized
by the decay rate $Dq^2+1/\tau_s$. However, the spin rotation mixes
the in-plane and out-of-plane spin dynamics. As a result, the
evolution is usually described by three relaxation rates. There are
two special cases where the evolution can be described by two
relaxation rates. The first one is in the system where only one of
the Dresselhaus and Rashba terms is important so that
the two in-plane spin dynamics become identical.
The other is the spin injection/diffusion along
$(110)$ and $(1\bar{1}0)$ axises where the net effective magnetic
field $\bar{{\bf h}}$ is parallel to one of the characteristic
directions of the in-plane spin dynamics. Thus it can only mix the remaining
in-plane spin with the out-of-plane spin dynamics.
The TSG evolution of these cases has a double-exponential form,
\begin{equation}
\label{szt}
S_z(q,t)=S_z(q,0)(\lambda_+e^{-t/\tau_+}+\lambda_-e^{-t/\tau_-})\ ,
\end{equation}
with relaxation rates
\begin{equation}
  \label{eq:decay}
  \Gamma_{\pm}=\frac{1}{\tau_{\pm}}=Dq^2+{1\over 2}
  ({1\over\tau_{s 1}}+{1\over\tau_s})
  \pm{1\over 2\tau_{s 2}}
  \sqrt{1+\frac{16Dq^2 \tau_{s 2}^2}{\tau^{\prime}_{s 1}}}\ ,
\end{equation}
in which
\begin{equation}
  \label{eq:lambda}
  \lambda_{\pm}={1\over 2}\left(1\pm
  {1\over\sqrt{1+16Dq^2\tau^2_{{s}2}/\tau^{\prime}_{{s}1}}}\right)\ .
\end{equation}
Here $\tau_{s1}$ ($\tau_{s2}$) is the spin relaxation time of the in-plane spin
which mixes (does not mix) with the out-plane spin due to the net
effective magnetic field.  For  spin injection/diffusion along $(110)$
axis, $\tau_{s1}= \langle(\alpha-\hat{\beta})^2
k^2\tau_1\rangle/2+\gamma ^2\langle k^6\tau_3\rangle/32$
and $\tau_{{s}1}^{\prime}=\langle(\alpha-\hat{\beta})^2
k^2\tau_1\rangle/2$.
For spin injection/diffusion along $(1\bar{1}0)$
axis, $\tau_{s1}= \langle(\alpha+\hat{\beta})^2
k^2\tau_1\rangle/2+\gamma ^2\langle k^6\tau_3\rangle/32$
and
$\tau_{{s}1}^{\prime}=\langle(\alpha+\hat{\beta})^2
k^2\tau_1\rangle/2$.
In the
long wave length limit ($q\ll 1$),
$\Gamma_+\simeq 1/\tau_s+(1+4\tau_{{s}2}/\tau_{{s}1}^{\prime})Dq^2$
and $\Gamma_-\simeq
1/\tau_{{s}1}+(1-4\tau_{{s}2}/\tau^{\prime}_{{s}1})Dq^2$
become quadratic functions of $q$,
roughly correspond to the out-of-plane and the in-plane relaxation rates
respectively.
In general both of these two decay rates
[Eq.\ (\ref{eq:decay})] are no longer simple
quadratic functions of $q$. If one uses the quadratic fitting
to yield the spin diffusion coefficient, one either gets
larger (for $\Gamma_+$) or smaller (for $\Gamma_-$) values than the
true spin diffusion coefficient. The accurate way to get the
information of spin diffusion coefficient should be from
the average of these two rates
\begin{equation}
\Gamma=(\Gamma_++\Gamma_-)/2=Dq^2+(1/\tau_s+1/\tau_{{s1}})/2\ ,
\label{eq:Gamma}
\end{equation}
which differs from
the current widely used formula by replacing the spin decay rate by the
average of the out-of-plane and in-plane ones.

\begin{figure}[htpb]
  \centering
   \epsfig{file=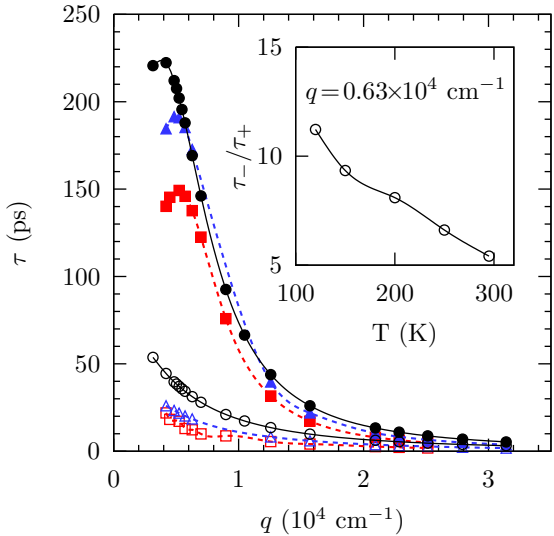,width=0.85\columnwidth}
   \caption{(Color online)  Spin relaxation times $\tau_{+}$
 (open) and $\tau_-$ (filled) {\em vs}. spin grating wave vector
 $q$ for three different temperatures $T=120$ (red boxes), 150
     (blue triangles) and 295\ K (black circles).
     The inset is the ratio $\tau_-/\tau_+$ as a function of temperature for
 $q=0.63\times 10^{4}$\ cm$^{-1}$.
   }
   \label{fig:tau}
 \end{figure}

\subsection{Numerical Results}
The spin diffusion coefficient obtained by the simplified kinetic
spin Bloch equations does not include the contribution of the spin
Coulomb drag since the electron-electron Coulomb scattering is
neglected.  Moreover, the simplified equations are derived with only
the elastic scattering. In the equations all of the relaxation times
$\tau_1$ that appear in the diffusion coefficient $D$, net
effective magnetic field $\bar{{\bf h}}$ as well as the spin
relaxation matrix are the same.
For the inelastic electron-phonon scattering,
it is still possible to write down Eq.\ (\ref{eq:2}) with same
relaxation time $\tau_1$ by the elastic scattering approximation.
This approximation is valid only for the electron-acoustic phonon
scattering at high temperature and is not valid for the electron-LO
phonon scattering. When the Coulomb scattering is important,
it gives different contributions to these relaxation times.
On the one hand, the relaxation time $\tau_1$ in the spin relaxation
matrix is affected by the whole Coulomb scattering. On the other
hand, since the Coulomb scattering among the same spin specie does not
change the motion of the center of mass, it does not directly affect the
spin diffusion coefficient. Therefore the relaxation time $\tau_1$
in the spin diffusion coefficient is only affected by  part of the Coulomb
scattering. For $\tau_1$ in the net effective magnetic field, it
is even more complicated to analyze the role of the Coulomb scattering
since it is the joint result of diffusion and spin
precession. Therefore the relaxation times $\tau_1$ in the three different
terms should be different in the present of
the  Coulomb scattering. It would be extremely difficult to get the
analytical results when the Coulomb scattering is taken into account.
In order to study the drag effect,
we numerically solve the full
kinetic spin Bloch equations Eq.\ (\ref{eq:kinetic}), with all the
scattering explicitly included. The numerical scheme is laid out in
detail in Appendix\ B. For simplicity, we first consider the spin
diffusion along (100) axis in a symmetrical QW in which the Rashba
term vanishes. In our calculation,
the SOC strength,
the electron and impurity densities,
 and the QW width are chosen to be
$\gamma=11.4$\ eV$\AA^3$,\cite{zhou_prb_2007}
$N=7.8\times
10^{11}$\ cm$^{-2}$, $N_i=1\times 10^{11}$\ cm$^{-2}$ and $a=12$\ nm
respectively. The material parameters are listed in detail in Refs.\
\onlinecite{weng_prb_2004b}.

Our numerical results
  show that the temporal evolution of TSG can not be fitted by a
  simple exponential function with desirable accuracy. However, if we
  use double-exponential function, the accuracy can be improved more
  than one order of magnitude. This justifies
that the temporal evolution of the TSG
has indeed double-exponential form in high temperature regime.
In Fig.\ 1 
we present the
relaxation times of TSG as a function of the grating wave-vector $q$
for temperatures $T=120,150$ and 295~K. It is seen that
$\tau_{+}$ decreases monotonically as $q$ increases while $\tau_{-}$
has a peak at some small $q_0$. The wave-vector of the peak
red-shifts when the temperature increases. In the
inset of Fig.\ 1, 
we show the ratio
$\tau_-/\tau_+$ as the function of temperature for fixed $q$. One finds
 that the ratio decreases with temperature.
Our results are in contrast to the predictions of earlier
theoretical works that the ratio of these two decay rates and the
position of the peak depend only on the SOC and material parameters,
but do not vary with the temperature.\cite{burkov_2004,bernevig_06}
Recent experiment showed that the ratio of these two decay rates are
indeed decreases with the increase of the temperature.\cite{weber_07} It
is understood that the temperature dependence of the peak position
and ratio between two decay times originate from the cubic $k$-term in
the Dresselhaus effective magnetic field. In earlier works, it was assumed that
only the linear term is important. However, in wide quantum
wells with high electron density, the cubic term becomes important.
Once the cubic term is considered, the peak $q_0$ moves from
$q_0=\sqrt{15}m\beta/2$ (which is independent of temperature) to about
$q_0=\sqrt{15}m\hat{\beta}/2=\sqrt{15}m(\beta-\gamma k^2/4){/2}$. Since
$\langle k^2\rangle$ increases with temperature, thus $q_0$
decreases. The temperature dependence of $\tau_+/\tau_-$ also
originates from the contribution of the cubic Dresselhaus term.

\begin{figure}[htbp]
  \centering
  \epsfig{file=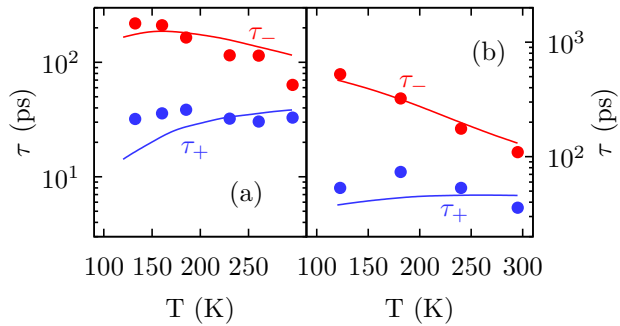,width=0.95\columnwidth}
   \caption{(Color online) Spin relaxation times $\tau_\pm$
{\it vs.} temperature
  for (a) high-mobility sample with $q=0.58\times 10^4$\ cm$^{-1}$
     and (b) low-mobility sample with $q=0.69\times
     10^4$\ cm$^{-1}$. The dots are the experiment data from
     Ref.\ \onlinecite{weber_07}.
   }
   \label{fig:fitting}
 \end{figure}

Due to the natural of the numerical calculation, it is not possible to
show the results with all possible parameter combinations. However our
qualitative conclusions are valid not just for this particular
parameter set but for a large range of parameters. In order to check
the quantitative accuracy of our numerical calculation we further show the
spin relaxation times as functions of temperature together with the
experimental data from Ref.\ \onlinecite{weber_07} in
Fig.\ 2.
Curves in  Fig.\ 2(a) 
are the theoretical spin relaxation times of high mobility
($\mu=1.5\times 10^4$\ cm$^2$/Vs) sample with $q=0.58\times
10^4$\ cm$^{-1}$, while the dots are the experimental data from
Ref.\ \onlinecite{weber_07}. Figure\ 2(b) 
shows the
result of low mobility sample ($\mu=3.5\times 10^3$\ cm$^2$/Vs) with
$q=0.69\times 10^4$\ cm$^{-1}$. In the calculation we use the finite
square well assumption.\cite{zhou_prb_2007} All the parameters we
use are chosen to be the experimental value if available, {\it eg.}
the grating wave-vector, the electron density, the quantum well width, and the
impurity concentration determined from the mobility. The only
adjustable parameters are the spin-orbit coupling coefficients
$\gamma$ and $\alpha$. In the calculation, $\gamma$ is chosen to be
$11.4$\ eV$\AA^3$ and $13.8$\ eV$\AA^3$ for the high and low mobility samples
respectively and $\alpha$ is set to be $0.3\beta$, close to the choice
in  Ref.\ \onlinecite{weber_07}. One can see from the figure that
our theoretical results are in fairly good agreement with the experiment data.

\begin{figure}[htpb]
   \centering
   \epsfig{file=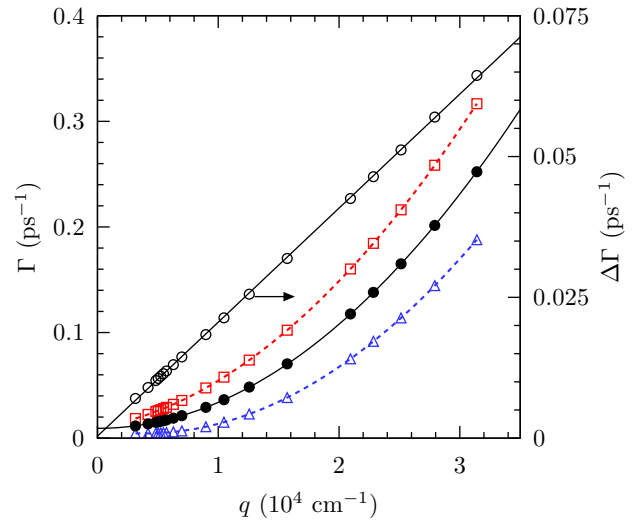,width=0.95\columnwidth}
   \caption{(Color online) $\Gamma=(\Gamma_++\Gamma_-)/2$ and
   $\Delta\Gamma=(\Gamma_+-\Gamma_-)/2$
 {\em vs}. $q$ at  $T=295$\ K.
     Open boxes/triangles are the relaxation rates $\Gamma_{+/-}$
 calculated from the full kinetic spin Bloch equations.
  Filled/open circles represent $\Gamma$ and $\Delta\Gamma$ respectively.
   Noted that the scale for $\Delta\Gamma$ is on the
   right hand side of the frame. The solid curves are the fitting to
     $\Gamma$ and $\Delta\Gamma$ respectively. The dashed curves are
   guide to eyes.}
   \label{fig:rate}
 \end{figure}

In Fig.\ 3,
we plot the decay rates
$\Gamma_{\pm}=1/\tau_{\pm}$ and their average
$\Gamma=(\Gamma_++\Gamma_-)/2$ and difference
$\Delta\Gamma=(\Gamma_+-\Gamma_-)/2$
as functions of $q$ at  $T=295$\ K.
The decay rates $\Gamma_{\pm}$ fit poorly with a quadratic function of
$q$. In contrast, the average decay rate $\Gamma$ fits pretty
well by the function $\Gamma=D_sq^2+1/\tau_s^{\prime}$.
The resident error of the quadratic
fitting for $\Gamma$ is two orders of magnitude smaller than
those of $\Gamma_{\pm}$. Moreover we find that
$\tau_s^{\prime}$ is very close to $4\tau_s/3$,
inverse of the
average of the in-plane and out-of-plane spin relaxation rate.
For example,  at $T=295$\ K,
$\tau_s^{\prime}$ is about $107.8$\ ps
compares to $4\tau_s/3=111.6$\ ps with
$\tau_s$ calculated by solving the
kinetic spin Bloch equations for spacial uniform system with the
same parameters.\cite{weng_prb_2003}
Inspired by Eq.\ (\ref{eq:Gamma}),
the coefficient of the quadratic term $D_s$ can be
reasonably assumed to be the spin diffusion coefficient. In this way
one can calculate the
spin diffusion coefficient with the effect of Coulomb drag included.
The difference of $\Gamma_{+}$ and $\Gamma_{-}$ fits well as a
linear function of $q$. The linear coefficient of $\Delta\Gamma$ is
about $2\sqrt{D_s/\tau^{\prime}_{{s}1}}$ from Eq.~(\ref{eq:decay}).

\begin{figure}[htpb]
   \centering
   \epsfig{file=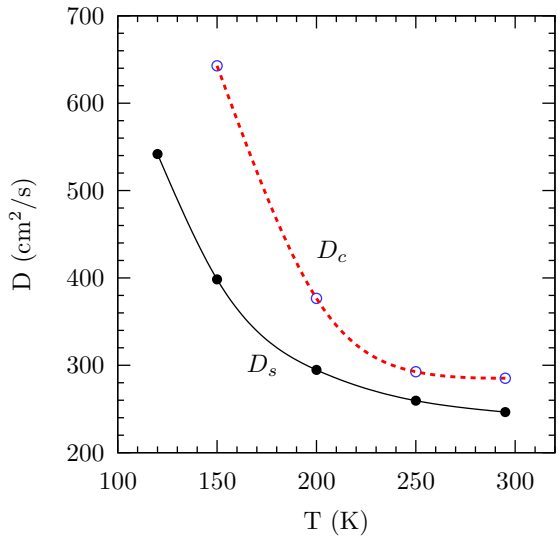,width=0.85\columnwidth}
   \caption{(Color online) Diffusion coefficient as a function of temperature.
   Solid circles: Spin diffusion constant $D_s$ with Coulomb drag;
     Open circles: Charge diffusion constant $D_c$.
   }
   \label{fig:DS}
 \end{figure}

In Fig.\ 4 
we present the spin diffusion coefficient
calculated in the above
mentioned method as a function of temperature. For comparison, we also
include the charge diffusion coefficient, which is calculated by solving the kinetic spin Bloch equations
with the initial condition being the charge gradient instead of the
spin gradient.
It is clearly seen from the figure that $D_s< D_c$.
For spin-unpolarized charge diffusion, the electrons move
along the same direction and the Coulomb scattering does not change the
center-of-mass motion, therefore it does not change the charge diffusion coefficient directly. However,
in the spin-polarized transport, spin-up and -down electrons move against
each other and the Coulomb scattering therefore slows down the
relative motion of these two spin species and reduces spin diffusion coefficient.
This is the so-called spin Coulomb drag effect.\cite{amico_2002,amico_2003,jiang_2005} From the figure one
can tell that, in the temperature regime we study,
as the temperature increases, both spin and charge diffusion
coefficients decrease and their difference also decreases. Therefore
the Coulomb drag is stronger in the low temperature regime. However,
even at room temperature the Coulomb drag is still strong enough to
reduce the diffusion coefficient by 30\ \%.
These results quantitatively agree with those of
Refs.\ \onlinecite{amico_2002,amico_2003}.

It is also pointed out
that the reduction of spin diffusion coefficient mostly comes from
the Coulomb drag. The SOC only has slightly effect on the diffusion
coefficient since the SOC is very small compared to the Fermi energy.
The numerical result shows that removing the SOC only changes
spin diffusion coefficient up to
one tenth percent for the system we studied.

\section{Steady-State  Spin Injection}
\label{statictransport}

In this section we discuss how to obtain the steady-state
spin injection information from the TSG signal, i.e.,
to find out the  relation between the steady-state spin injection
length $L_s$ and  spatial spin oscillation ``period'' $L_0$  and
the spin diffusion constant $D_s$ together with the spin
relaxation time $\tau_s$.

We first show from the simplified solution presented
in Sec.\ \ref{simp} that the steady-state spin injection can be extracted from
the  TSG signal by integrating the  TSG signal
Eq.\ (\ref{szt})
over the time from 0 to $\infty$ and the wave-vector from $-\infty$ to
$\infty$.  
Form Eqs.\ (\ref{eq:decay}) and (\ref{eq:lambda}), the integrated TSG reads
$S_z(x)=S_z(0)e^{-x/L_s}\cos (x/L_0+\psi)$  with
\begin{eqnarray}
  \label{eq:L_s} L_s&=&\sqrt[4]{D^2_s\tau_{{s}1}\tau_s} /\sin{\phi\over 2}\ ,\\
  \label{eq:L_0} L_0&=&\sqrt[4]{D^2_s\tau_{{s}1}\tau_s}/\cos{\phi\over 2}\ .
\end{eqnarray}
In these equations
\begin{equation}
  \label{eq:angle}
  \cos\phi=\sqrt{\tau_{{s}1}\tau_s}
(4/\tau^{\prime}_{{s1}}-1/\tau_{{s}1}-1/\tau_s)/2.
\end{equation}
It is noted that
if one only considers the Rashba term or the linear Dresselhaus term,
$\tau^{\prime}_{{s}1}=\tau_{{s}1}$, one
then recovers  the result
$L_s=2\sqrt{D_s\tau_s}$ from linear response theory.\cite{burkov_2004}
It is seen that the spin precession actually prolongs the out-of-plane
spin injection length by mixing the fast decay of the out-of-plane
spin with  the slow decay of the in-plane spin. It is further noted
from Eq.\ (\ref{eq:L_s}) that {\em the spin
injection length $L_s$ is generally larger than
$\sqrt{D_s\tau_s}$}. The only exception is when the spin injection is along
$(110)$-direction for the QW with  equal linear
Dresselhaus and Rashba spin-orbit couplings.
In this case $\theta=\pi/4$, therefore the net effective magnetic field
 [Eq.\ (\ref{eq:hq})] vanishes. Consequently the in-plane and
out-of-plane spin modes do not mix.

\begin{table*}[htpb]
  \centering
   \begin{tabular}{cccccccccc}\hline\hline
 $\alpha$  & {$\mbox{Injection} \atop \mbox{direction}$} &
 $D_s$\ (cm$^{2}$/s) & $\tau_s^{\prime}$\ (ps) &
 $c$\ ($\mu$m/ps) & $d$\ (ps$^{-1}$) & $L_s^{{\mathtt{T}}}$\ ($\mu$m) &
 $L_0^{\mathtt{T}}$\ ($\mu$m)
 & $L_s^{\mathtt{S}}$\ ($\mu$m) & $L_0^{\mathtt{S}}$\ ($\mu$m) \\
 \hline
 $\alpha=0$ & ($100$) & 246 & 107.8 & 0.02 & $3.5\times 10^{-4}$ & 2.27 & 2.44 &
 2.28 & 2.46 \\
 \hline
 $\alpha=\beta$ & (110) & 243 & 49.7 & $9.4\times 10^{-3}$ & 0.012 & 0.92 & 5.1 &
 1 & 4.9 \\
 \hline
 $\alpha=\beta$ & ($1\bar{1}0$) & 250 & 28.1 & 0.058 & $-2.7\times 10^{-3}$ & 2.7 & 0.86 &
 2.1 & 0.9\\
\hline\hline
   \end{tabular}
 \caption{Comparison of spin injection length $L_s$ and spatial oscillation
  length $L_0$ along different injection directions at $T=295$\ K
 from different approaches. The superscripts ``S'' and ``T''
 stand for $L_{s}$ and $L_0$ obtained from the injection calculation
in the steady state and those from TSG
  parameters by using Eqs.\ (\ref{ls}) and (\ref{l0}) respectively. }
  \label{tab:parms}
 \end{table*}

For most of the cases, $L_s$ is in the same order of $\sqrt{D_s\tau_s}$,
although the former is usually larger. However,
there are some special cases where $\phi\rightarrow 0$, $L_s$ can be {\em
orders of magnitude} different from $\sqrt{D_s\tau_s}$.
Specifically, according to Cheng {\em et al.},\cite{cheng_prb_2007}
when the spin  injection/transport direction
is  along $(1\bar{1}0)$ in (001) QW, $L_s$ becomes larger and larger
as $\alpha$ approaches $\beta$, {\em regardless} of the direction
of spin polarization.
At the limit of $\alpha=\beta$, 
the spin injection length  trends to infinity when the cubic
Dresselhaus term is ignored and the spin oscillates
with a spatial period of $2\pi/(2m\beta)$.\cite{cheng_prb_2007}
As $\tau_s$ is finite and $D_s$ changes little for different
grating directions, there is no way to obtain infinite $L_s$ from
$\sqrt{D_s\tau_s}$.

The infinite injection length and the finite oscillation period can also be
understood from the TSG point of view.
Without the cubic Dresselhaus term,
$1/\tau_{s2}=(\alpha-\beta)^2\langle k^2\tau_1\rangle/2$
approaches zero and
$1/\tau_{s1}\rightarrow 1/\tau_s=2\beta^2\langle k^2\tau_1\rangle$
when $\alpha\rightarrow\beta$. Consequently
$\tau_{\pm}=(\sqrt{D}q\pm 1/\sqrt{\tau_s})^{-2}$ when $\alpha=\beta$. It
is then straightforward to see that $\tau_-$
becomes infinite  provided
$q=q_0=1/\sqrt{D\tau_s}=2m\beta$.
Therefore the steady-state spin injection
along $(1\bar{1}0)$ axis is dominated by this non-decay
TSG mode which is responsible for the infinite spin injection length and the
spatial oscillation period $2\pi/q_0=2\pi/(2m\beta)$.

When the electron-phonon and the electron-electron Coulomb scatterings are
taken into account,  $\phi$ should be revised accordingly.
Unfortunately, there are no  analytical expressions for
$\tau_s$, $\tau_{s1}$ and $\tau_{s1}^{\prime}$ when all the
scatterings are included. Nevertheless, one can directly use the TSG result
to calculate the static injection parameters.
The numerical result indicates that the two decay rates obey
\begin{equation}
\Gamma_{\pm}=D_sq^2+1/\tau^{\prime}_s\pm(cq+d)
\end{equation}
where $c$ and $d$ are the fitting parameters to
\begin{equation}
\Delta\Gamma=cq+d\ .
\end{equation}
Again by integrating the TSG signal over the time and wave-vector $q$
with above fitted $\Gamma_\pm$, one obtains
\begin{eqnarray}
\label{ls}
L_s&=&2D_s/\sqrt{|c^2-4D_s(1/\tau^{\prime}_s-d)|}\ ,\\
L_0&=&2D_s/c\ .
\label{l0}
\end{eqnarray}
We stress that these two equations give the right spin injection length and
the spin oscillation period in the presence of the SOC.
From the experiment point of view, one can monitor the
time evolution of TSG with different wave-vectors $q$ and obtain the
corresponding decay rates $\Gamma_\pm$.
From the $q$-dependence of the decay rates,
one can  calculate the spin injection length and spin oscillation
period from Eqs.\ (\ref{ls}) and (\ref{l0}).
It is noted that
Eq.\ (\ref{ls}) naturally gives the infinite injection length in the special case\cite{cheng_prb_2007}
discussed above as $c=2\sqrt{D_s/\tau_s}$ and $d=0$ so that the
denominator in Eq.\ (\ref{ls}) tends to zero.
In contrast, $\sqrt{D_s\tau_s}$ always remains finite.

In order to check that the accuracy of this approach [Eqs.\ (\ref{ls}) and
(\ref{l0})], we
compare $L_s$ and $L_0$ obtained from directly numerically solving the kinetic spin Bloch
equations [Eq.\ (\ref{eq:kinetic})]
for spin injection in the steady state as described in
Refs.\ \onlinecite{cheng_jap_2007,cheng_prb_2007},
 with those  from  the TSG
approach. In Table\ \ref{tab:parms} we list $L_s$ and
$L_0$ obtained from the TSG signal and from
the steady-state solution of the kinetic
spin Bloch equations  for spin injection direction along
(100) axis in symmetrical (001) QW, as well as (110) and ($1\bar{1}0$) axes in
asymmetrical (001) QW with equal Dresselhaus and Rashba coupling.
For the sake of clarity,
we also list the corresponding fitted parameters
$\tau_s^\prime$, $c$ and $d$ in the table. It is noted that in
the calculation  the cubic Dresselhaus term is included. One can see
from the table that even though the injection and the oscillation lengths
at different conditions are quite different, the spin diffusion coefficients
$D_s$  are almost the same. Their differences are within
the numerical fitting error. This confirms that the SOC is too small
compared to the Fermi energy to affect
$D_s$. Moreover, for all the cases we study, $L_0$ and $L_s$ obtained from
these two methods agree with each other very well.
Although we should point out that the accuracy of $L_0$ is higher than $L_s$
 due to the numerical error analysis.
Since $c^2-4D_s(1/\tau_s^{\prime}-d)$ is close to 0, its
numerical error is much larger than $c$. Therefore
$L_s$ calculated from TSG is not as accurate
as $L_0$.

\section{Conclusion}

In conclusion, we study the evolution of TSG by solving the kinetic spin Bloch
equations with periodic boundary condition.
From the simplified  equations, we analytically show
that due to the spin precession the spin signal decays
double-exponentially instead of simple exponentially, even in the diffusive
regime. {\em Neither} of these two decay rates are simple quadratic
functions of the grating wave-vector $q$. However their average value depends
quadratically on $q$ and the corresponding coefficient of the
quadratic term is the right diffusion coefficient $D_s$. Therefore it is
more accurate to yield the diffusion coefficient from $q$ dependence
of the average of the two decay rates.

We further show that the corresponding solution from
 the full kinetic spin Bloch
equations which include all of the scattering mechanisms, especially
the Coulomb scattering, is also in the form of the double exponential
decay. From the $q$-dependence of the average decay rate, one can
calculate the spin diffusion coefficient with the Coulomb drag effect
included. It is shown
that the Coulomb drag effect is stronger in low temperature and
decreases with the increase of  temperature. However, even at room
temperature the Coulomb drag is still an important factor which
reduces the spin diffusion coefficient markedly compared to
 the charge diffusion coefficient.

We also show that by using the TSG result one can obtain
the  characteristic steady-state transport
parameters, such as the injection length
and the spatial oscillation length, from diffusion coefficient $D_s$
and the spin relaxation times. We point out that in a system
with the SOC, the wildly adopted relation
$L_s=\sqrt{D_s\tau_s}$ in the literature is
generally quantitatively inaccurate  and can be even
qualitatively wrong for some special cases.
 The accurate way to extract the spin injection
 and spatial oscillation lengths directly from the
TSG decay rates at different grating wave-vector, which are experimentally
 measurable, is proposed.  We believe this investigation is important
in fully understanding the  TSG signals in
experiment.

\begin{acknowledgments}
This work was supported by the Natural Science Foundation of China
under Grant Nos.\ 10574120 and 10725417, the National Basic Research
Program of China under Grant No.\ 2006CB922005, the Knowledge
Innovation Project of Chinese Academy of Sciences  and the US Army
Research Office. The authors would like to thank T. Korn for his
critical reading of this manuscript. One of the authors (MWW)
would like to thank J. Fabin and
C. Sch\"uller at Universit\"at Regensburg, Germany for hospitality
where this work
was finalized, and the Robert-Bosch Stiftung and GRK 638 for financial support.

\end{acknowledgments}

\appendix
\section{Derivation of Simplified Solution}

By using a similar method for calculating the spin relaxation due to
the Dresselhaus effect, Eqs.~(\ref{eq:2}) and (\ref{eq:hq}) can be derived
by expanding Eq.~(\ref{eq:kinetic}) in angular momentums.\cite{dp,dpb}
Neglecting the Hartree-Fock term, the inelastic scattering and the
electric field, the Fourier component of the $l$-th order of density
matrix obeys the following equation:
\begin{widetext}
\begin{eqnarray}
&&  {\partial \rho_{l}(q,k,t)\over\partial t}-
{ikq\over 2m}\biggl(\rho_{l+1}(q,k,t)+\rho_{l-1}(q,k,t)\biggr)
-i\sum_m\biggl[{\bf h}_{l-m}(k)\cdot
\mbox{\boldmath $\sigma$},\rho_{m}(q,k,t)\biggr]\nonumber
\\ &&
=-{\rho_{l}(q,k,t)\over\tau_l}(1-\delta_{l,0}),
\label{eq:klth}
\end{eqnarray}
where $\rho_l(q,k,t)=\int
e^{-il\phi}\rho_{{\bf k}}(q,t) {d\phi/2\pi}$ and
${\bf k}=(k\cos\phi,k\sin\phi,0)$.
When both the Dresselhaus and the Rashba
terms are taken into account,
\begin{eqnarray}
{\bf h}({\bf k})&=&
\gamma k\bigl((\pi/a)^2-k^2/4\bigr)
(-\cos\phi,\sin\phi,0)+
\gamma k^3/4(-\cos 3\phi,-\sin 3\phi,0)\nonumber\\ &&
+\alpha k(\sin\phi,-\cos\phi,0),\\
&=&\hat{\beta} k
(-\cos\phi,\sin\phi,0)
+\alpha k(\sin\phi,-\cos\phi,0)
+\gamma k^3/4(-\cos 3\phi,-\sin 3\phi,0).
\end{eqnarray}
\end{widetext}
Therefore there are four effective magnetic field components which
do not vanish, namely ${\bf h}_{\pm 1}(k)$ and ${\bf h}_{\pm 3}(k)$.
It should be noted that, in quasi-two-dimensional system, the $\pm
1$ components of the Dresselhaus term are  modified by the cubic
term $\hat{\beta}=\gamma[(\pi/a)^2-k^2/4]=\beta-\gamma k^2/4$. When
the scattering is strong, one can drop the terms with $\tau_l$
higher than the first order and rearrange Eq.\ (\ref{eq:klth}) to
obtain Eq.\ (\ref{eq:2}). The only difference is that in the spatial
inhomogeneous system, we have additional terms relying on the
wave-vector $q$. These additional terms give rise to
the second and third terms in Eq.\ (3).

\section{Numerical  Scheme}

In order to solve the kinetic spin Bloch equations numerically, one has to
discretize the real space, the momentum space as well as the  time. The real
space is divided into segments with equal length. The momentum space
is divided into  grids of equal energy and angular
differences.\cite{weng_prb_2004b}
The second order up-wind differential scheme is applied for the
 diffusion term and the drift term. The former reads
\begin{equation}
  {k_x\over m}{\partial f(x)  \over\partial x}
  \rightarrow \left\{
    \begin{array}{cc}
      {k_x\over m}{3f(x)-4f(x-\Delta x)+f(x-2\Delta x)\over 2\Delta
        x} & {k_x\over m}>0 \\
      -{k_x\over m}{3f(x)-4f(x+\Delta x)+f(x+2\Delta x)\over 2\Delta
        x} & {k_x\over m}<0
    \end{array}\right.\ .
\end{equation}
The boundary condition for $x$ is chosen to be the periodic one when we
calculate the TSG problem or fixed when we calculate the steady-state
injection problem.\cite{cheng_jap_2007}
In the energy (${\cal E}$) and angular ($\phi$) space, the drift term reads
\begin{eqnarray}
eE(x){\partial g(k_x,k_y)\over\partial k_x}&=&{eE(x)\over \sqrt{2m}}\biggl
(\frac{\partial}{\partial{\cal E}}[{2\sqrt{\cal E}\cos\theta
g({\cal E},\theta)}]\nonumber\\
&&-\frac{\partial}{\partial \theta}\frac{ g({\cal E},
\theta)\sin\theta }{\sqrt{\cal E}
}\biggr)\ .
\end{eqnarray}
Similar to the diffusion term, one can easily write down the second
order up-wind differential schemes for ${\cal E}$ and $\theta$
respectively. The numerical schemes for the spin precession and the
scattering terms are laid out in detail in
Refs.\ \onlinecite{weng_prb_2004b,cheng_jap_2007}.

We apply the
third order semi-implicit Adams-Bashforth scheme for the time
differential\cite{ab3} to achieve higher accuracy in temporal
evolution. This scheme also saves CPU time. The differential
scheme is then given by
\begin{eqnarray}
&&{\rho_{{\bf k}}(x,t+\Delta t)-\rho_{{\bf k}}(x,t)\over\Delta t}=
{5\over 4}F[\rho_{{\bf k}}(x,t+\Delta t)]\nonumber\\
&&\hspace{0.5cm}
-F[\rho_{{\bf k}}(x,t)]
+3/4F[\rho_{{\bf k}}(x,t-\Delta t)]+
{23\over 12}G[\rho_{{\bf k}}(x,t)]\nonumber\\
&&\hspace{0.5cm}
-{16\over 12}G[\rho_{{\bf k}}(x,t-\Delta t)]
+{5\over 12}G[\rho_{{\bf k}}(x,t-2\Delta t)]\ .
\end{eqnarray}
Here $F[\rho_{{\bf k}}(x,t)]$
denotes the drift and diffusion terms (the second and third terms)
in Eq.\ (\ref{eq:kinetic})
and $G[\rho_{{\bf k}}(x,t)]$ stands for the spin precession and the
scattering terms (the fourth and fifth terms)
in Eq.\ (\ref{eq:kinetic}).
The implicit part of the equation is solved by
Jacobian-free Newton-Krylov algorithm.\cite{brown_1990}

The accuracy of the numerical scheme used in this paper is higher
than the one used in our previous
works.\cite{weng_jap_2003,cheng_jap_2007} The main numerical
errors come from the drift and diffusion parts since the grid size
of real space and momentum space is limited by the computing power.
We find that the accuracy of the temporal evolution does
 not change the result of the
final steady-state
spin transport too much. It is therefore expected that the present
numerical scheme and the previous one give very close results on the
steady-state transport properties. However, the present scheme enables us to
also study the time sensitive phenomenons such as the
 TSG to a sufficient accuracy.


\end{document}